\def\year{2020}\relax
\documentclass[letterpaper]{article} 

\usepackage{mdframed} 
\usepackage{xcolor} 
\usepackage{mathrsfs} 
\newcommand{\DDE}{\ensuremath{\mathcal{{DDE}}}}
\newcommand{\DCEC}{\ensuremath{{\mathcal{{DCEC}}}}}

\usepackage{amsmath}

\newcommand{\lsort}[1]{%
  \ensuremath{\mbox{\textsf{#1}}}}
\newcommand{\defsort}[2]{%
  \newcommand{#1}{\lsort{#2}}}
\defsort{\Action}{Action}
\defsort{\Time}{Time}
\defsort{\Self}{Self}

\defsort{\Agent}{Agent}
\defsort{\Entrant}{Entrant}
\defsort{\ActionType}{ActionType}
\defsort{\Moment}{Moment}
\defsort{\Boolean}{Formula}
\defsort{\Fluent}{Fluent}
\defsort{\Event}{Event}
\defsort{\Object}{Object}
\defsort{\RealTerm}{RealTerm}

\defsort{\Numeric}{Numeric}
\defsort{\Number}{Number}

\newcommand{\lsymbol}[1]{%
  \ensuremath{\mathit{#1}}}
\newcommand{\defsymbol}[2]{%
  \newcommand{#1}{\lsymbol{#2}}}
\defsymbol{\action}{action}
\defsymbol{\initially}{initially}
\defsymbol{\holds}{Holds}
\defsymbol{\happens}{happens}
\defsymbol{\clipped}{clipped}
\defsymbol{\initiates}{initiates}
\defsymbol{\terminates}{terminates}
\defsymbol{\prior}{prior}
\defsymbol{\interval}{interval}

\newcommand{\lconstant}[1]{%
  \ensuremath{\mbox{\textsf{#1}}}}
\newcommand{\defconstant}[2]{%
  \newcommand{#1}{\lconstant{#2}}}
\defconstant{\Block}{Block}
\defconstant{\Prev}{Prev}

\newcommand{\lmodality}[1]{%
  \ensuremath{\mathbf{#1}}}
\newcommand{\defmodality}[2]{%
  \newcommand{#1}{\lmodality{#2}}}
\defmodality{\common}{C}
\defmodality{\knows}{K}
\defmodality{\believes}{B}
\defmodality{\perceives}{P}

\defmodality{\desires}{D}
\defmodality{\intends}{I}
\defmodality{\says}{S}
\defmodality{\ought}{O}

\usepackage{aaai20}  

\usepackage{times}  
\usepackage{helvet} 
\usepackage{courier}  
\usepackage[hyphens]{url}  
\usepackage{graphicx} 
\usepackage{graphicx}  
\title{AI Can Stop Mass Shootings, and More}
\author{Selmer Bringsjord $\bullet$ Naveen Sundar Govindarajulu
$\bullet$ Michael Giancola\\
Rensselaer AI \& Reasoning (RAIR) Lab\\
Department of Cognitive Science; Department of Computer Science\\
Rensselaer Polytechnic Institute (RPI);
Troy NY 12180 USA\\
selmer.bringsjord@gmail.com $\bullet$ naveensundarg@gmail.com $\bullet$ mike.j.giancola@gmail.com}

\begin{document}
\maketitle

\begin{abstract}
  We propose to build directly upon our longstanding, prior r\&d in
  AI/machine ethics in order to attempt to make real the blue-sky idea
  of AI that can thwart mass shootings, by bringing to bear its
  ethical reasoning.  The r\&d in question is overtly and avowedly
  logicist in form, and since we are hardly the only ones who have
  established a firm foundation in the attempt to imbue AI's with
  their own ethical sensibility, the pursuit of our proposal by those
  in different methodological camps should, we believe, be considered
  as well.  We seek herein to make our vision at least somewhat
  concrete by anchoring our exposition to two simulations, one in
  which the AI saves the lives of innocents by locking out a
  malevolent human's gun, and a second in which this malevolent agent
  is allowed by the AI to be neutralized by law enforcement.  Along
  the way, some objections are anticipated, and rebutted.
\end{abstract}

\section{Introduction}
\label{sect:intro}

No one reading this sentence is unaware of tragic mass shootings in
the past.  Can future carnage of this kind be forestalled?  If so,
how?  Many politicians of all stripes confidently answer the first
question in the affirmative, but unfortunately then a cacophony of
competing answers to the second quickly ensues.  We too are optimistic
about tomorrow, but the rationale we offer for our sanguinity has
nothing to do with debates about background checks and banning
particular high-powered weapons or magazines, nor with a hope that the
evil and/or insane in our species can somehow be put in a kind of
perpetual non-kinetic quarantine, separated from firearms.  While we
hope that such measures, which of late have thankfully been gaining
some traction, will be put in place, our optimism is instead rooted in
AI; specifically, in \emph{ethically correct} AI; and even more
specifically still: our hope is in ethically correct AI that guards
guns.  Unless AI is harnessed in the manner we recommend, it seems
inevitable that politicians (at least in the U.S.)~will continue to
battle each other, and it does not strike us as irrational to hold
that even if some legislation emerges from their debates, which of
late seems more likely, it will not prevent what can also be seen as a
source of the problem in many cases: namely, that guns themselves have
no ethical compass.


\section{What Could Have Been}
\label{sect:what_could_have_been}

A rather depressing fact about the human condition is that any number
of real-life tragedies in the past could be cited in order to make our
point regarding what could have been instead; that is, there have been
many avoidable mass shootings, in which a human deploys one or more
guns that are neither intelligent nor ethically correct, and innocents
die or are maimed.  Without loss of generality, we ask the reader to
recall the recent El Paso shooting in Texas.  If the kind of AI we
seek had been in place, history would have been very different in this
case.  To grasp this, let's turn back the clock.  The shooter is
driving to Walmart, an assault rifle, and a massive amount of
ammunition, in his vehicle.  The AI we envisage knows that this weapon
is there, and that it can be used only for very specific purposes, in
very specific environments (and of course it knows what those purposes
and environments are).  At Walmart itself, in the parking lot, any
attempt on the part of the would-be assailant to use his weapon, or
even position it for use in any way, will result in it being locked
out by the AI.  In the particular case at hand, the AI knows that
killing anyone with the gun, except perhaps e.g.\ for self-defense
purposes, is unethical.  Since the AI rules out self-defense, the gun
is rendered useless, and locked out.  This is depicted pictorially in
Figure \ref{fig1}.

Continuing with what could have been: Texas Rangers were earlier
notified by AI, and now arrive on the scene.  If the malevolent human
persists in an attempt to kill/maim despite the neutralization of his
rifle, say be resorting to a knife, the Rangers are ethically cleared
to shoot in order to save lives: their guns, while also guarded by AI
that makes sure firing them is ethically permissible, are fully
operative because the Doctrine of Double Effect (or a variant; these
doctrines are discussed below) says that it's ethically permissible to
save the lives of innocent bystanders by killing the criminal.  They
do so, and the situation is secure; see the illustration in
Figure \ref{fig2}.  Unfortunately, what we have just described is an
alternate timeline that did not happen --- but in the future, in
similar situations, we believe it could, and we urge people to at
least contemplate whether we are right, and whether, if we are, such
AI is worth seeking.

\section{Can This Blue-Sky AI Really be Engineered?}
\label{sect:can_be_engineered}

Predictably, some will object as follows: ``The \emph{concept} you
introduce is attractive.  But unfortunately it's nothing more than a
dream; actually, nothing more than a \emph{pipe} dream.  Is this AI
really feasible, science- and engineering-wise?''  We answer in the
affirmative, confidently.  The overarching reason for our optimism is
that for well over 15 years Bringsjord and colleagues have been
developing logicist AI technology to install in artificial agents so
as to ensure that these agents are ethically correct
[e.g.\ \cite{sb_etal_ieee_robots,aaai_fall05symp_machethics,sb_divine_command_robots,bello_topoi_moral_machine,nsg_sb_dde_ijcai}].
This research program has reached a higher degree of maturity during a
phase over the past six years, during which the second author,
Govindarajulu, has collaborated with Bringsjord, and led on many
fronts, including not only papers that seek to formalize and implement
ethical theories in AIs
[e.g.\ \cite{nsg_sb_dde_ijcai,toward_engineering_virtuous_machines}],
but also in the development of high-powered automated reasoning
technology ideal for machine ethics; for instance the automated
reasoner ShadowProver \cite{shadowprover,DBLP:journals/corr/abs-1912-12959}, and the planner
Spectra \cite{spectra}, which is itself built up from automated
reasoning.

Importantly, while all of the longstanding work pointed to in the
previous paragraph is logicist, and thus in line with arguments in
favor of such AI
[e.g.\ \cite{logicist_manifesto,do_machine-learning_machines_learn}],
we wish to point out that other work designed to imbue AIs with their
own ethical reasoning and decision-making capacity is of a type that
in our judgment fits well our logicist orientation
[e.g.\ \cite{arkin_governing_lethal_book,programming_machine_ethics_pereira}],
and with our blue-sky vision.  But beyond this, since of course lives
are at stake, we call for an ecumenical outlook; hence if
statistical/connectionist ML can somehow be integrated with
transparent, rigorous ethical theories, codes, and principles [and in
fact some guidance for those who might wish to do just this is
provided in \cite{nsg_sb_dde_ijcai}] that can serve as a verifiable,
surveyable basis for locking out weapons, we would be thrilled.

\section{Why is Killing Wrong?}
\label{sect:why_killing_wrong}

As professional ethicists know, it's rather challenging to say why
it's wrong to kill people, especially if one is attempting to answer
this question on the basis of any consequentialist ethical theory
(e.g.\ utilitarianism); a classic, cogent statement of the problem is
provided in \cite{whats_wrong_with_killing_people}.  We are inclined
to affirm the general answer to the first question in the present
section's title that runs like this: ``To kill a human person $h$
is \textit{ipso facto} to cut off any chance that $h$ can reach any of
the future goals that $h$ has.  This is what makes killing an innocent
person intrinsically wrong.''  This answer, formalized, undergirds the
first of our two simulations.

\section{Automating the Doctrine of Double Effect}
\label{sect:doctrine_nable_effect}

We referred above to the Doctrine of Double Effect, $\DDE$ for
short.
We now informally but rigorously present this ethical principle, so
that the present short paper is self-contained.  Our presentation
presupposes that we possess an ethical hierarchy that classifies
actions (e.g.\ as \textit{forbidden}, \textit{morally
neutral}, \textit{obligatory}); see \cite{ethical_hierarchy_icre2015}.
We further assume that we have a utility or goodness function for
states of the world or effects; this assumption is roughly in line
with a part of all consequentialist ethical theories (e.g.\
utilitarianism).  For an autonomous agent $a$, an action $\alpha$ in a
situation $\sigma$ at time $t$ is said to be
$\DDE$-compliant \textit{iff}:

\begin{small}
\begin{quote}
\begin{enumerate}
\item[$\mathbf{C}_1$] the action is not forbidden (where we assume an
  ethical hierarchy such as the one given by Bringsjord
  \shortcite{ethical_hierarchy_icre2015}, and require that the action be
  neutral or above neutral in such a hierarchy);
\item[$\mathbf{C}_2$]  the net utility or goodness of the action is greater than some positive
  amount $\gamma$;
\item[$\mathbf{C}_{3a}$] the agent performing the action intends only the good effects;
\item[$\mathbf{C}_{3b}$] the agent does not intend any of the bad effects;
\item[$\mathbf{C}_4$] the bad effects are not used as a means to
  obtain the good effects; and
\item[$\mathbf{C}_5$] if there are bad effects, the agent would rather
  the situation be different and the agent not have to perform the
  action. That is, the action is unavoidable.
\end{enumerate}
\end{quote}
\end{small}

See Clause 6 of Principle III in \cite{khatchadourian1988principle}
for a justification of clause $\mathbf{C}_5$.\footnote{This clause has
not been discussed in any prior rigorous treatments of \DDE, but we
feel $\mathbf{C}_5$ captures an important part of \DDE\ as it is
normally used, e.g.\ in unavoidable ethically thorny situations one
would rather not be present in.  $\mathbf{C}_5$ is necessary, as the
condition is subjunctive/counterfactual in nature and hence may not
always follow from $\mathbf{C}_1 - \mathbf{C}_4$, since there is no
subjunctive content in those conditions.  Note that
while \cite{pereira2016counterfactuals} model \DDE\ using
counterfactuals, they use counterfactuals to model $\mathbf{C}_4$
rather than $\mathbf{C}_5$.  That said, the formalization of
$\mathbf{C}_5$ is quite difficult, requiring the use of
computationally hard counterfactual and subjunctive reasoning.  We
leave this aside here, reserved for future work.}  Most importantly,
note that \DDE\ has long been taken as the ethical basis for
self-defense, and just war \cite{sep_DDE}.  Our work brings this
tradition, which has been informal, into the realm of formal methods,
and our second simulation is based upon an AI proving that
\DDE\ holds.

\section{Two Simulations}
\label{sect:2_simulations}

\begin{figure}[hbt]
\centering
\includegraphics[width=0.8\columnwidth]{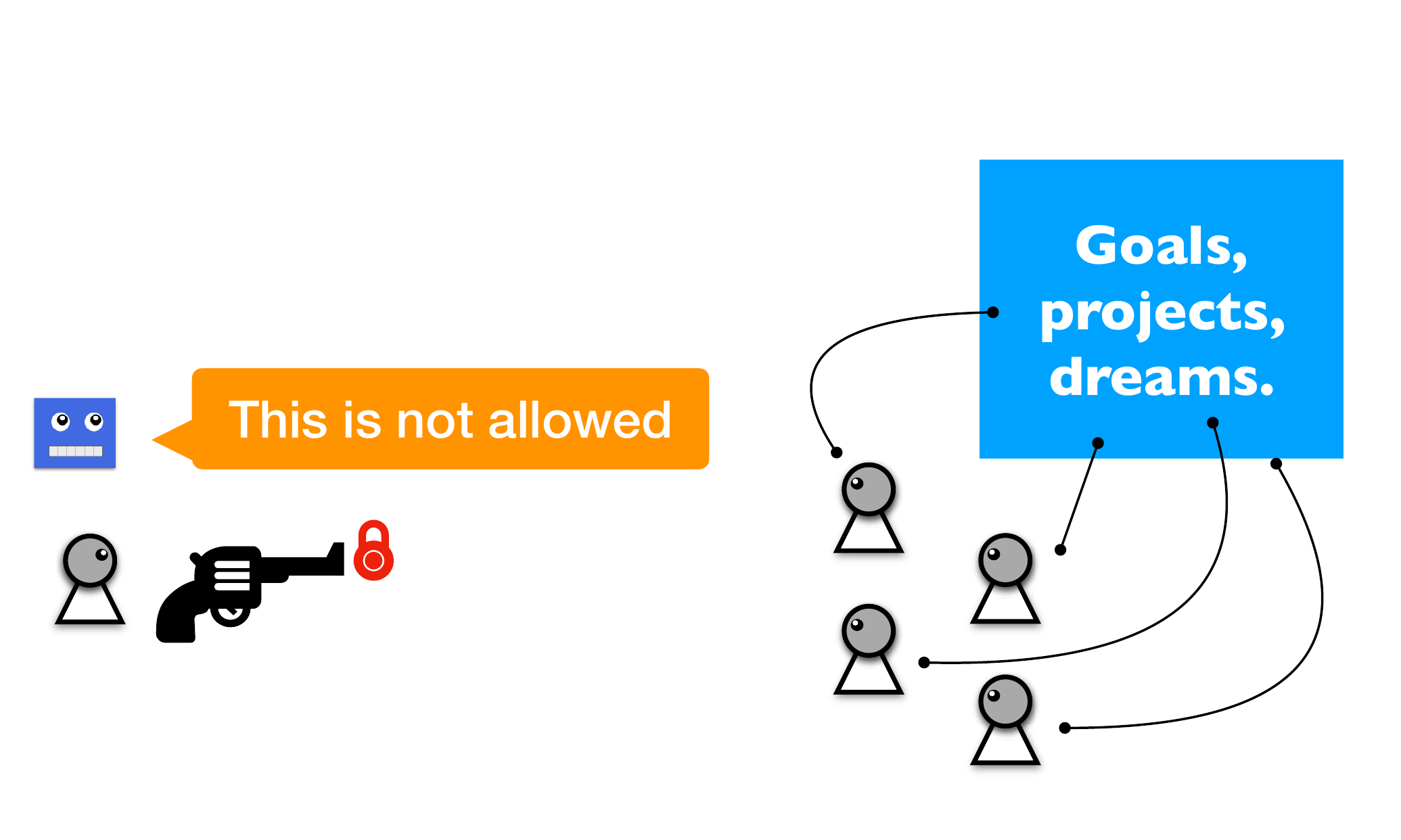} 
\caption{Prohibition Against Killing in Force; AI Thwarts Malevolent Assailant.  \textit{This corresponds to Simulation 1.}}
\label{fig1}
\end{figure}

\begin{figure}[hbt]
\centering
\includegraphics[width=0.8\columnwidth]{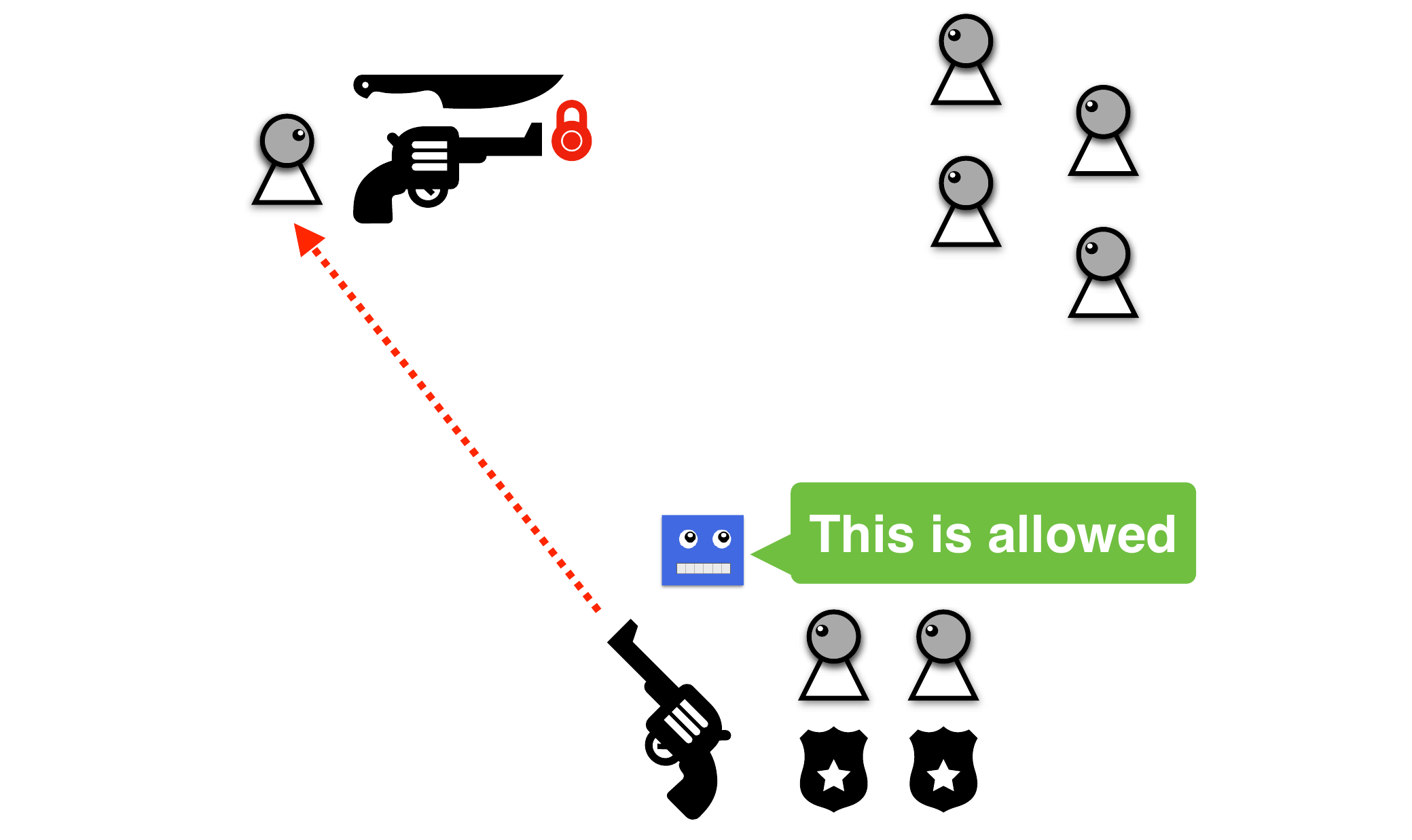} 
\caption{$\mathcal{DDE}$ Sanctions Shooting Malevolent Assailant; AI Refrains from Thwarting.  \textit{This corresponds to Simulation 2.}}
\label{fig2}
\end{figure}

A pair of simulations, each confessedly simple, nonetheless lend
credence to our claim that our blue-sky conception is feasible.  In
the first, an AI blocks the pivotal human action $\alpha$ because the
action is (given, of course, a background ethical theory that is
presumed) ethically impermissible.  Essentially, the AI is able to
prove $\mathbf{O}(a, \neg \alpha)$ by using a principle of the form
$\Phi \rightarrow \mathbf{O}(a, \neg \alpha)$.  Here $\Phi$ says that
performance of $\alpha$ by $a$ would deprive an innocent person $a'$
of the ability to continue to pursue, after this deprivation, any of
his/her goals.  Once the AI, powered by ShadowProver, proves that
$\alpha$ is ethically impermissible for $a$, an inability to prove by
\DDE\ that there is an ``override'' entails in this simulation that the
pivotal action cannot be performed by the human.  In the second
simulation, the AI allows a human action by \DDE\ that directly kills
one (the malevolent shooter) to save four human members of law
enforcement (see Fig.\ \ref{fig1}).  Here now is a brutally brief look
on the more technical side of the simulations in question.

As discussed earlier, it is difficult to state exactly why it's
intrinsically wrong to kill people.  Yet we must do exactly this if we
are to enable a machine to generate a proof (or even just a cogent
argument) that the assailant's gun should, on ethical grounds, be
locked.  Moreover, we must state this as formulae expressed in a
formal logic that an automated theorem prover can reason over.  In our
case, we utilize the Deontic Cognitive Event Calculus (\DCEC) and the
aforementioned ShadowProver, respectively.  Much has been written
elsewhere about \DCEC\ and the class of calculi that subsumes it;
these details are out of scope here, and we direct interested readers
to \cite{nsg_sb_dde_ijcai}, which makes a nice starting place for
those in AI.  The original cognitive calculus appeared long ago,
in \cite{ka_sb_scc_seqcalc}; but this calculus had no ethical
dimension in the form of deontic operators, and pre-dated ShadowProver
[and used Athena instead, a still-vibrant system that anchors the
recent \cite{fpmics}].  Here it should be sufficient to say only that
dialects of \DCEC\ have been used to formalize and automate highly
intensional reasoning processes, such as the false-belief task
\cite{ka_sb_scc_seqcalc} and \textit{akrasia} (succumbing to temptation to
violate moral principles) \cite{akratic_robots_ieee}.  \DCEC\ is a
  sorted (i.e.\ typed) quantified multi-operator modal logic.  The
  calculus has a well-defined syntax and proof calculus; the latter is
  based on natural deduction
\cite{gentzen_investigations_into_logical_deduction}, and includes all
the introduction and elimination rules for second-order logic, as well
as inference schemata for the modal operators and related structures.
The modal operators in \DCEC\ include the standard ones for knowledge
$\knows$, belief $\believes$, desire $\desires$, intention $\intends$,
and in some dialects operators for perception and communication as
well.  The general format of an intensional operator is e.g.\
$\knows\left(a, t, \phi\right)$, which says that agent $a$ knows at
time $t$ the proposition $\phi$.  Here $\phi$ can in turn be any
arbitrary formula.

As to the pair of simulations themselves, while a full discussion of
them would not fit within the limitations of this short paper, we do
discuss one critical definition next, that of the (abstracted)
predicate $\Prev(x,y,g,a,t)$, which means that $x$ prevents $y$ from
achieving goal $g$ via action $a$ at time $t$; in a form expressed
in \DCEC\ syntax:


\begin{tiny}
\begin{equation*}
\begin{aligned}
&\exists t_1, t2: \Moment \\
&\wedge\left\{
\begin{aligned}
  &\prior(t, t_1),\\
  &\prior(t_1, t_2),\\
  &\knows\Big(x, t, \desires(y, t, \holds(g, t_2)) \wedge \intends(y, t, \happens(g, t_2))\Big),\\
  &\knows\left(x, t, \exists a': \ActionType \left(
      \begin{aligned} &\intends(y, t_1, \happens(\action(y, a'), t_1)) \\
& \ \ \ \ \ \ \ \ \ \ \ \ \ \ \ \ \ \ \ \ \wedge \\
        &\left[\begin{aligned}
            &\left(\begin{aligned}&\happens(\action(y,a'),t_1) \\ 
                                  & \ \ \ \ \ \ \ \ \ \ \ \ \ \ \wedge \\
                                  &\lnot \Block(x,y,g,a,t)\end{aligned}\right) \\
              &\ \ \ \ \ \ \ \ \ \ \rightarrow \happens(g, t_2) \\
          \end{aligned}\right]
      \end{aligned}\right)\right),\\
  &\knows\Big(x, t, \happens(\action(x, a), t) \rightarrow \Block(x,y,g,a,t)\Big), \\
  &\happens(\action(x,a),t)
\end{aligned}\right\}\\
\end{aligned}
\end{equation*}
\end{tiny}

The key components in this definition are:

\begin{quote}
\begin{enumerate}
\item $x$ knows that $y$ desires a goal $g$ and intends to accomplish $g$;
\item $x$ knows that $y$ intends to perform an action $a'$ that will lead to the
accomplishment of $y$'s goal $g$, unless $x$ does something to block
that goal;
\item $x$ knows that if $x$ performs action $a$ then $y$'s goal $g$ will be blocked; and
\item $x$ performs $a$.
\end{enumerate}
\end{quote}

Utilizing this definition, along with a few other formulae in \DCEC\
(chiefly, that preventing another human from achieving their goals,
unless overridden by \DDE, is forbidden), ShadowProver can prove ---
on an Apple laptop, and without any human-engineered optimization ---
for Simulation 1 that lock-out must happen in less than a second, and 3
seconds for Simulation 2 that lock-out must not happen.

\section{Why Not \emph{Legally} Correct AIs Instead?}
\label{sect:why_not_legal}

We expect some readers to sensibly ask why we don't restrict the AI we
seek to \emph{legal} correctness, instead of ethical correctness.
After all (as it will be said), the shootings in question are illegal.
The answer is that, one, much of our work on the deontic-logic side
conforms to a framework that Leibniz espoused, in which legal
obligations are the ``weakest'' kind of moral
obligations/prohibitions, and come just before, but connected to,
ethical obligations in the hierarchy $\mathscr{EH}$, first introduced
in \cite{ethical_hierarchy_icre2015}.  In this Leibnizian approach,
there is no hard-and-fast breakage between legal
obligations/prohibitions and moral ones; the underlying logic is
seamless across the two spheres.  Hence, any and all of our formalisms
and technology can be used directly in a ``law-only'' manner.  This is
in fact provably the case; some relevant theorems appear
in \cite{ethical_hierarchy_icre2015}.  The second part of our reply to
the present objection is that we wish to ensure that AIs can be
ethically correct even in cases where the local laws are wildly
divergent from standard Occidental ethical theories.

\section{Additional Objections}
\label{sect:objections}

Of course, there are any number of additional objections that will be
raised against the research direction we seek to catalyze by the
present short paper.  It is fairly easy to anticipate many of them,
but current space constraints preclude presenting them, and then
providing rebuttals.  We rest content with a speedy treatment of but
two objections, the first of which is:

  \begin{quote} 
    
    ``Consider the \textit{Charlie Hebdo} tragedy, in Paris.  Here,
    high-powered rifles were legally purchased in Slovakia, modified,
    and then smuggled into France, where they were then horribly
    unleashed upon innocent journalists.  Even if the major gun
    manufacturers, like the major car manufacturers, willingly subject
    themselves to the requirement that their products are infused with
    ethically correct AI of the type you are engineering, surely there
    will still be `outlaw' manufacturers that elude any AI aboard
    their weapons.''

  \end{quote}

In reply, we note that our blue-sky conception is in no way restricted
to the idea that the guarding AI is only in the weapons in question.
Turn back the clock to the \textit{Hebdo} tragedy, and assume for the
sake of argument that the brothers' rifles in question are devoid of
any overseeing AI of the type present in the two simulations described
above.  It still remains true, for example, that the terrorists in
this case must travel to Rue Nicolas-Appert with their weapons, and
there would in general be any number of options available to AIs that
perceive the brothers in transit with their illegal cargo to thwart
such transit.  Ethically correct AI, with the power to guard human
life on the basis of suitable ethical theory/ies, ethical codes, and
legal theory/ies/codes, deployed in and across a sensor-rich city like
Paris, would have any number of actions available to it by which a
violent future can be avoided in favor of life.  Whether guarding AI
is in weapons or outside them looking on, certain core requirements
must be met in order to ensure efficacy.  For instance, here are two
(put roughly) things that a guarding AI should be able to come to
know/believe:

\begin{footnotesize}
\begin{mdframed}[linecolor=white,frametitle=Epistemic Requirements for Weapon-Guarding AI,frametitlebackgroundcolor=gray!25,
                 backgroundcolor=gray!10, 
                 roundcorner=8pt]

  Given any human $h$, at any point of time $t$, an ethically correct,
  overseeing AI should \emph{at least} be able to come to know/believe
  the following, in order to verify that relevant actions on the part
  of $h$ are $\mathcal{DDE}$-compliant (where $\phi$ is a
  state-of-affairs that includes use of a weapon).

\begin{enumerate}
  \item The human's intentions: $(\lnot)\mathbf{I}\left(h, t, \phi\right)$
  \item Forbiddenness/Permissibility: $(\lnot)\mathbf{O}\left(a, t, \sigma, \lnot\phi\right)$
\end{enumerate}
\end{mdframed}
\end{footnotesize}

Now here is the second objection:

  \begin{quote} 
    
    ``Your hope for AI will be dashed by the brute fact that AI in
    weapons can be discarded by hackers.''

  \end{quote}

\noindent
This is an objection that we have long anticipated in our work devoted
to installing ethical controls in such things as robots, and we see no
reason why our approach there, which is to bring machine ethics down
to an immutable hardware level
\cite{moral_robots_op_sys_level,ethical_operating_systems}, cannot be
pursued for weapons as well.  Of course, a longer discussion of the
very real challenge here is needed.

\section{Concluding Remarks}
\label{sect:conclusion}

Alert readers may ask why the ``, And More'' appears in our title.
The phrase is there because machine ethics, once one is willing to
look to AI itself for moral correctness, and protective actions
flowing therefrom, can be infused in other artifacts the full
``AI-absent'' human control of which often results in carnage.  A
classic example is driving.  We all know that AI has made amazing
strides in self-driving vehicles, but there is no need to wait for
lives to be saved by broad implementation of self-driving AI:
ethically correct AI, today, can shut down a car if the would-be human
driver is perceived by an artificial agent to be intoxicated (above,
say, .08 BAC).  In 2017 alone, over 10,000 people died in the U.S.\
because of intoxicated human drivers used their vehicles
immorally/illegally
(NHTSA\footnote{\url{https://crashstats.nhtsa.dot.gov/Api/Public/ViewPublication/812630}}).
Ethically correct AI, indeed relatively such AI, can stop this, today.

We end with a simple observation, and from it a single question: Many
researchers are already working on the challenge of bringing ethically
correct AIs to the world.  Why not channel some of this ingenious work
specifically into the engineering of AIs that are employed to guard
artifacts that, indisputably, are all too often vehicles for unethical
agents of the human sort to cause horrible harm?

\section{Acknowledgments}

The authors are indebted to ONR for a 6-year MURI grant devoted to the
science and engineering of morally competent AI/robots (M.\ Scheutz
PI, Co-PIs S.\ Bringsjord \& B.\ Malle, N.S.\ Govindarajulu Senior
Research Scientist), focused in our case on the use of multi-operator
quantified modal logics to specify and implement such competence; and
to AFOSR (S.\ Bringsjord PI) for support that continues to enable the
invention and implementation of unprecedentedly expressive
computational logics and automated reasoners that in turn enable
human-level computational intelligence.

\clearpage
\bibliographystyle{aaai}
\bibliography{main72,naveen}

\end{document}